\documentclass[12pt,preprint]{aastex}

\begin{document}
\title{The Origin and Motion of PSR J0538+2817 in S147}
\author{C.-Y. Ng\altaffilmark{1}, Roger W. Romani\altaffilmark{1}, Walter F. Brisken\altaffilmark{2}, Shami Chatterjee\altaffilmark{2,3}, \& Michael Kramer\altaffilmark{4}}
\altaffiltext{1}{Department of Physics, Stanford University, Stanford, CA 94305.}
\altaffiltext{2}{National Radio Astronomy Observatory, P.O. Box O, Socorro, NM 87801.}
\altaffiltext{3}{Harvard-Smithsonian Center for Astrophysics, 60 Garden Street, MS 67, Cambridge, MA 02138.}
\altaffiltext{4}{Jodrell Bank Observatory, University of Manchester, Macclesfield, Cheshire SK11 9DL, UK.}
\email{ncy@astro.stanford.edu, rwr@astro.stanford.edu, wbrisken@aoc.nrao.edu, schatterjee@cfa.harvard.edu, mkramer@jb.man.ac.uk}

\begin{abstract}
	We report on VLBA astrometry and {\it CXO} imaging of PSR J0538+2817
in the supernova remnant S147. We measure a parallax distance of 
$1.47^{+0.42}_{-0.27}\;$kpc along with a high-precision proper motion,
giving a transverse velocity $V_\perp=400^{+114}_{-73}\;\mathrm{km\;s^{-1}}$.
A small extended wind nebula is detected around the pulsar; the symmetry axis
of this structure suggests that the spin axis lies $12\arcdeg\pm4\arcdeg$ from
the velocity vector (2-D), but the emission is too faint for robust model
independent statements. The neutron star is hot, consistent with the young
$\sim 40$\,kyr kinematic age. The pulsar progenitor is likely a runaway
from a nearby cluster, with NGC 1960 (M36) a leading candidate.
\end{abstract}
                                                                                

\section{INTRODUCTION}
\begin{figure}[hp]
\begin{center}
\includegraphics[angle=270,scale=0.8]{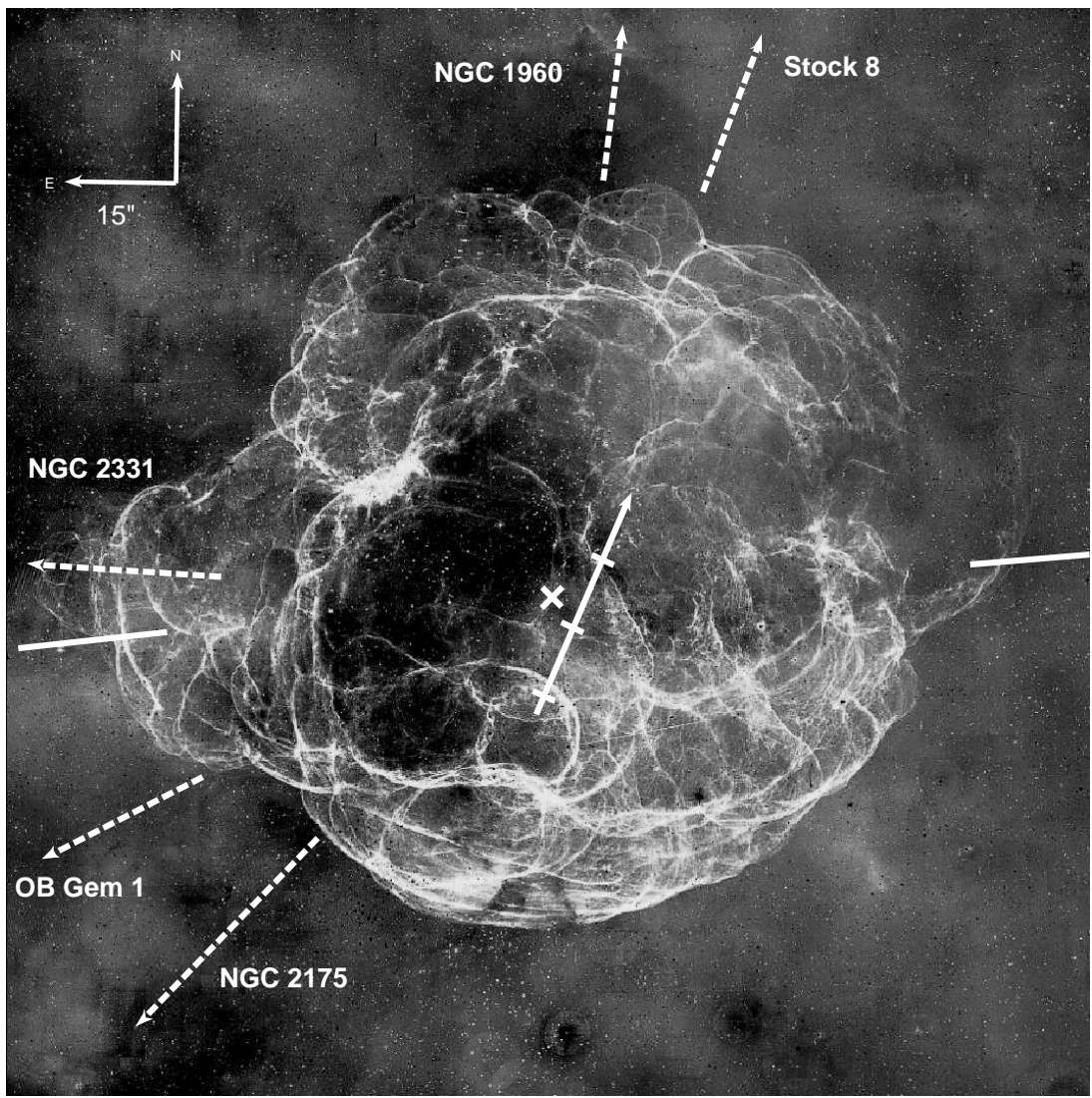}
\caption{Continuum-subtracted H$\alpha$ image of S147 \citep{dre05}.
The arrow shows the pulsar's proper motion direction and points to
its current position (see \S~\ref{s2}), with tick marks indicating
the birth-sites of the pulsar if born 20, 40 \& 60~kyr ago. The cross marks 
the SNR geometrical center suggested by \citet{kra03}. Solid lines
mark a possible bilateral axis \citep{gva06} and dashed lines
show the direction to several candidate birth-sites for the progenitor star.
\label{f1}}
\end{center}
\end{figure}

\object{S147} (G180.0$-$1.7) is an optically faint shell-type supernova
remnant (SNR) located in the direction of the Galactic anti-center. 
It is highly filamentary and has a radius of 83\arcmin\ \citep{sof80}.
Although age estimates vary widely, from 20~kyr \citep{sof80} to 
100~kyr \citep{kun80}, S147 is believed to be one of the oldest SNRs with
well-defined shell structure. It has been extensively studied at several
wavelengths including radio \citep{kun80,sof80},
optical \citep{loz76, kir79}, UV \citep{phi81}
and X-ray \citep{sou90}. A recent continuum-subtracted H$\alpha$
image \citep{dre05}
reveals much detailed structure in S147.
As shown in Fig.~\ref{f1}, the shell is mostly circular with a few obvious
blowouts to the East and the North. \citet{gva06} suggested that a faint
blowout to the West and the brighter lobe to the East define a
bilateral axis passing through the center.
Most of the bright filaments concentrate on the southern half,
where the SNR boundary is also sharper. In contrast, the northern half
is more diffuse and less well-defined.

Within the boundaries of S147, a 143~ms pulsar \object{PSR J0538+2817}
was discovered by \citet{and96} in an undirected pulsar survey
using the 305~m Arecibo radio telescope. The pulsar has a large
characteristic age $\tau_c = P/2\dot{P} = 620\;$kyr and a dispersion measure
(DM)-estimated distance of 1.2~kpc. Based on the positions and
distances of PSR J0538+2817 and S147, \citet{and96} suggested
an association between the two. From the maximum SNR age of $10^5\;$yr,
\citet{rom03} argued that PSR J0538+2817 had a slow initial spin
period. This was supported by \citet{kra03} who obtained a
rough timing proper motion which indicated that the pulsar is moving from
near the SNR geometrical center with a kinematic age of only 30~kyr,
much shorter than the characteristic age. To reconcile the discrepancy,
they also suggested a large initial spin period of 139~ms for the pulsar. 
Alternatively, \citet{gva06} proposed that PSR~J0538+2817 arose in
the first supernova from a massive binary, while a second supernova 
produced S147.
In X-rays, PSR J0538+2817 has been detected by \textit{ROSAT}
All Sky Survey \citep{sun96} and HRI imaging. A 20~ks Chandra ACIS-S
observation discovered extended emission around the pulsar \citep{rom03},
while an XMM-Newton observation \citep{mcg03} reported
the detection of pulsed X-ray emission.

In summary, S147 with an old nearly circular shell and a central pulsar showing
extended X-ray emission, provides some unique opportunities to test the
details of the pulsar/SNR connection. Kinematic measurements, along with X-ray
morphology and cooling measurements can probe the history of the pulsar.

\section{PARALLAX AND PROPER MOTION MEASUREMENT}

PSR J0538+2817 was observed with the NRAO VLBA over 9 epochs between 2002 and 2006.
The observations were conducted between 1.4 and 1.7 GHz, as a trade-off between
increasing pulsar flux at lower frequencies and improved resolution as
well as reduced ionospheric effects at higher frequencies. At each epoch, 4
spectral windows (IFs) each with 8 MHz bandwidth are observed simultaneously.
To retain visibility phase coherence, repeated visits were made to the primary
phase calibrator source 133\arcmin\ distant from the pulsar.  
A VLA survey of the region around J0538$+$2817 revealed a compact 8~mJy
source only 8\arcmin\ distance from the pulsar; the two sources are within the
same VLBA primary beam and were observed simultaneously.
The data were correlated in two passes, first centered on the pulsar field,
and then a second pass centered on the in-beam calibrator.
The signal-to-noise ratio for the pulsar was boosted by `gating' (accepting
signal from) the correlator on at the expected times of arrival of pulses, using
current pulse timing solutions obtained for each epoch from ongoing
observations at the Jodrell Bank Observatory.  Incremental phase calibration 
derived from observations of the in-beam source were applied to the pulsar for
considerably improved calibration.

Data analysis was performed using the AIPS package with a customized pipeline
(Brisken 2007, in preparation), which included amplitude calibration based
on the system temperatures and antenna gains, followed by visibility phase
calibration based on the primary calibrator and the in-beam source. Large scale
ionospheric phase effects were estimated and corrected with the AIPS task TECOR,
using global models of the ionospheric electron content based on distributed GPS
measurements. Once the calibration was completed, the pulsar position was measured by
fitting a Gaussian ellipse to calibrated images made separately for each IF. 
The pulsar parallax and proper motion were then obtained from a linear
least-squares fit to the epoch positions.

Table~\ref{tab1} shows the best-fit astrometric results from 8 good epochs,
for which the pulsar is far away from the sun, with IFs equally weighted.
The derived pulsar distance and transverse velocity
are $d=1.47^{+0.42}_{-0.27}\;$kpc and
$V_\perp=400^{+114}_{-73}\;\mathrm{km\;s}^{-1}$ respectively. The distance
is in good agreement with the DM-estimate value
$d_{\mathrm DM}=1.2\pm0.2\;$kpc, using the NE2001 model
\citep{cor02}. The proper motion is also consistent with the timing
observation results reported by \citet{kra03}, but our errors are much
smaller, especially in the ecliptic latitude ($\mu_\beta$).\label{s2}

To convert the proper motion to its local standard of rest (LSR),
we correct for differential Galactic rotation using
$\Omega_0=220\;\mathrm{km\;s}^{-1}$, $R_0=8.5\;$kpc \citep{fic89}
and the solar constants $10\pm0.36$, $5.25\pm0.62$, $7.17\pm0.38\;
\mathrm{km\;s}^{-1}$ \citep{deh98}. The corrected proper motion
is $\mu_*=58.51\pm0.18\;\mathrm{mas\;yr^{-1}}$ at position angle (PA)
$336\fdg8\pm0\fdg13$, which converts to a transverse velocity of
$V_\perp=407^{+116}_{-74}\;\mathrm{km\;s}^{-1}$.
Backward extrapolation of the pulsar's motion indicates that it passed 
$\sim8\arcmin$ from the geometrical center of S147 defined by 
\citet{kra03} at $\alpha=\mathrm{05^h40^m01^s\pm2^s},\;
\delta=27\arcdeg48\arcmin09\arcsec\pm20\arcsec$ (J2000).
Since the pulsar is young and located near the Galactic plane, this 
trajectory is not significantly altered by acceleration in the Galactic 
potential. For any reasonable Galactic model \citep[e.g. in][and references therein]{sun04}, 
the displacement from a linear trajectory over the pulsar characteristic 
age is much smaller than 1\arcmin. However, the SNR and pulsar are
almost certainly associated (see below) and thus we conclude that the
nominal center of the shell does not represent the explosion site and
that S147 underwent asymmetric expansion, likely due to
inhomogeneities in the surrounding ISM.

\begin{deluxetable}{ll}
\tablecaption{PARAMETERS FOR PSR J0538+2817\label{tab1}}
\tablewidth{0pt}
\tablehead{\colhead{Parameter} & \colhead{Value} }
\startdata
Epoch (MJD) & 53258 \\
R.A. (J2000) & $\mathrm{05^h38^m25\fs05237\pm0\fs00001}$ \\
Decl. (J2000) & $+28\arcdeg17\arcmin09\farcs3030\pm0\farcs0001$ \\
$\mu_{\alpha\cos\delta}$ (mas yr$^{-1}$) & $-23.53\pm 0.16$ \\
$\mu_\delta$ (mas yr$^{-1}$) & $52.59\pm 0.13$ \\
$\pi$ (mas) & $0.68\pm0.15$ \\
$d$ (kpc) & $1.47^{+0.42}_{-0.27}$ \\
$V_\perp$ (km s$^{-1}$) & $400^{+114}_{-73}$ \\
$l,\;b$ & $179\fdg7186,\; -1\fdg6859$ \\
$\mu_l,\;\mu_b$ (mas yr$^{-1}$) & $-57.03$, 8.18 \\
$n_e$ (cm$^{-3}$) & $0.027\pm 0.006$ \\
\enddata
\end{deluxetable}

\section{CXO OBSERVATIONS}
Chandra observations of PSR J0538+2817 were carried out on 2006
Jan 18 \& 20 (Observation IDs \dataset [ADS/Sa.CXO#obs/06242] {6242} \&
\dataset [ADS/Sa.CXO#obs/05538] {5538}) with the ACIS-I array operating
in very faint timed exposure (VF TE) imaging mode. The pulsar was
positioned near the aim point on the I3 chip, which was the only chip active
during the observation. Subarray mode with 160 rows
was used to further reduce the CCD pile-up. The resulting frame time of
0.5\,s reduced the pile-up of the pulsar to 2\%, ensuring the distortion
of the PSF and the high energy spectrum is minimal.
The total live time was $93.2\;$ks and examination of the background light
curve showed no strong flares during the observation. Hence, all data is
included in the analysis.

For comparison, we have also reprocessed the archival 20~ks ACIS-S exposure
(\dataset [ADS/Sa.CXO#obs/02796] {ObsID 2796}) observed on 2002 Feb 7.
After filtering out the periods suffering from background flares,
18.5 ks of clean exposure remains. All data analysis was performed using
CIAO 3.3 and CALDB 3.2.1 to ensure the latest time-dependent gain
calibration and charge transfer inefficiency (CTI) correction are applied.
To further improve the spatial resolution, we removed the ACIS pixel
randomization and applied the algorithm by \citet{mor01} to correct the
position of split pixel events. 

\section{SPATIAL ANALYSIS}
\begin{figure}[!h]
\includegraphics[angle=270]{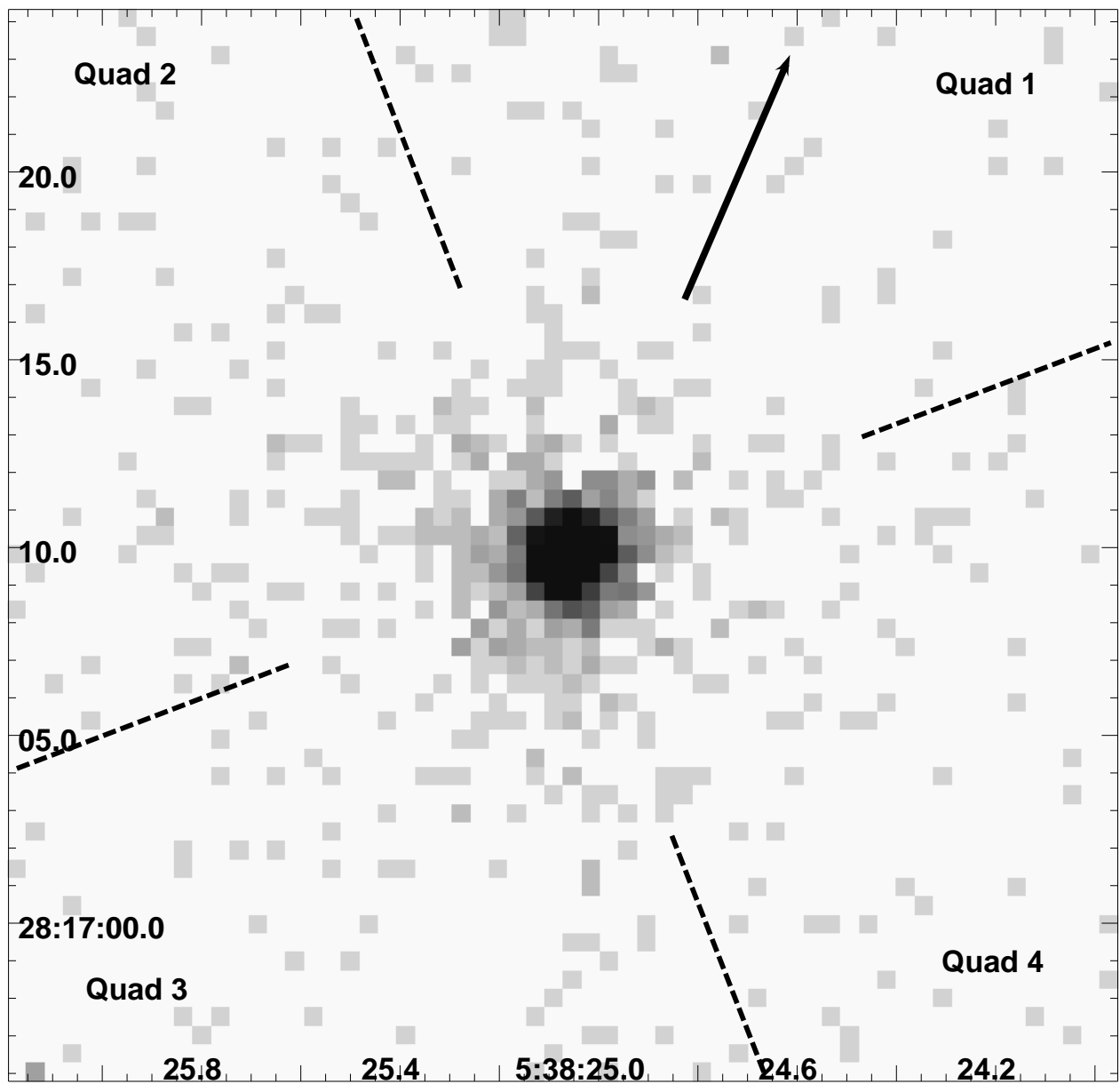}
\caption{ACIS-I 0.5-8 keV image. The arrow indicates the pulsar's proper
motion direction in its LSR. The dotted lines separate the four quadrants used to
measure the azimuthal distribution of the extended emission.\label{f2}}
\end{figure}

The ACIS-I 0.5-8 keV image is shown in Fig~\ref{f2}. The point source produces
$0.12\;\mathrm{cts\;s^{-1}}$. With the short frame time, a trail of counts 
can be seen along the readout direction (at PA~5$\arcdeg$), with
$0.5\;\mathrm{cts\;pixel^{-1}}$. The extended emission $\sim5\arcsec$
NE of the pulsar reported by \citet{rom03} is clearly detected in
the new observation. However, the deeper exposure shows that the structure
is not an obvious torus.  It also appears somewhat fainter in the new data, 
with $(4.3\pm0.7)\times 10^{-4}\;\mathrm{cts\;s^{-1}}$ in 0.5-8~keV after
background subtraction, as compared to $(9.7\pm2.2)\times 10^{-4}\;
\mathrm{cts\;s^{-1}}$ in the archival ACIS-S data. Diffuse emission is
also seen in the SE direction, as noted by \citet{rom03},
at $\sim3\arcsec$ from the point source. Given that the
direction is behind the pulsar's proper motion, this could be trailed
emission from relativistic electrons in the motion-confined pulsar wind,
as observed in other PWN systems.

\begin{figure}[!h]
\plotone{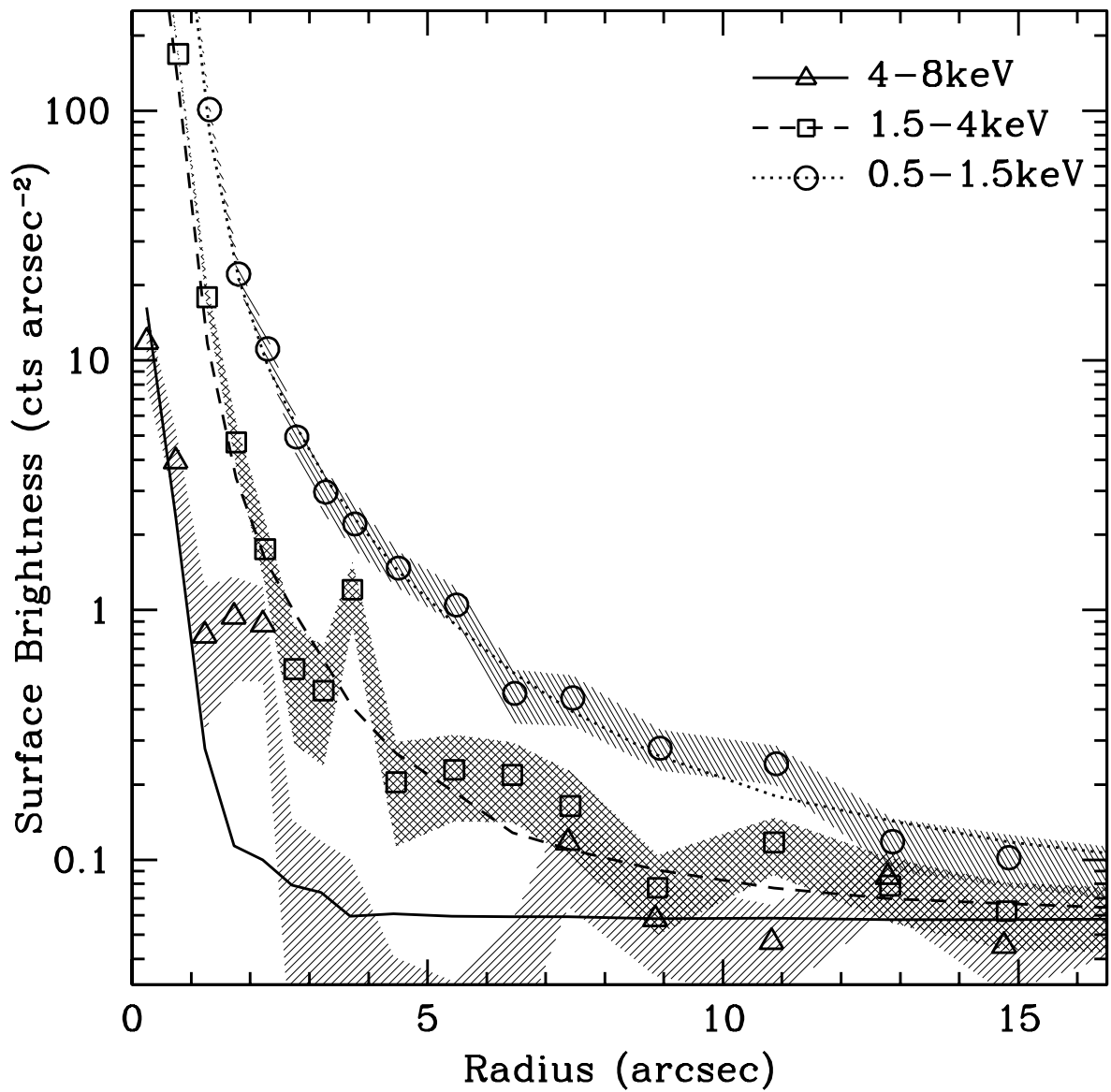}
\caption{Observed source (points) and model point source (line)
radial count distributions in three energy bands,
with the corresponding uncertainties (shaded bands).\label{f3}}
\end{figure}

Fig.~\ref{f3} compares the observed surface brightness of the source for the
ACIS-I data and the model PSF+background in 3 energy bands. The PSF is
simulated using the Chandra Ray-Tracer (ChaRT) and MARX software using
the best-fit pulsar spectrum from a 1\arcsec\ radius aperture, in order
to minimize any nebular contamination. Although the readout trail is
simulated in the model PSF, we excluded two rectangular regions of
$1\farcs5$ wide along the readout trail and beyond $2\farcs5$ from the
point source in our analysis, in order to improve the statistics. The graph
suggests that the data and PSF model are well matched for the low energy band,
while excess counts appear at several radii for the higher energies.

\begin{figure}[!h]
\begin{center}
\includegraphics[angle=270,scale=0.65]{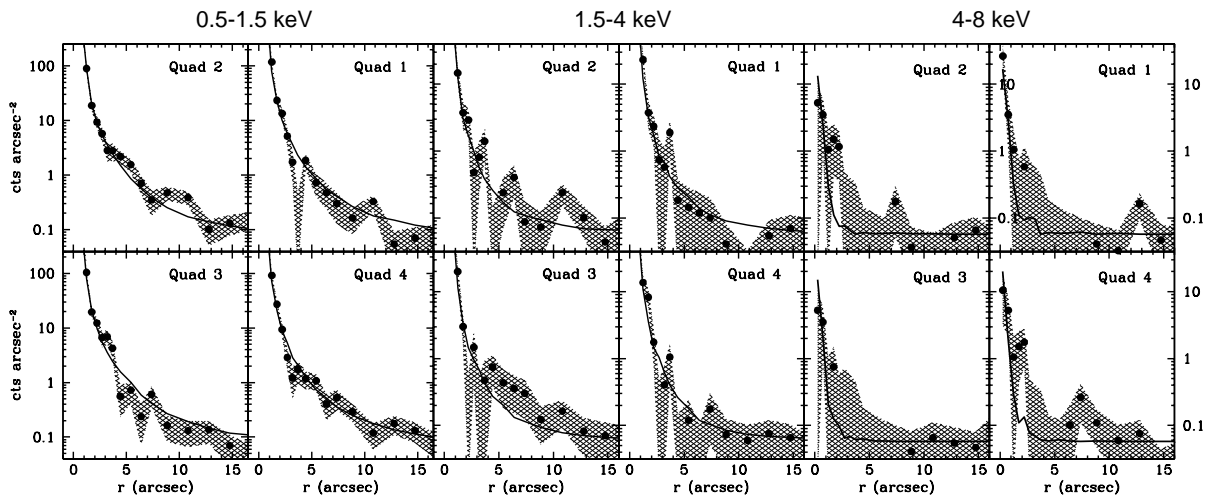}
\caption{Surface brightness of the observed source (points) and model
point source (lines) for the four quadrants in three energy bands.
The y-scale on the left applies to the 0.5-1.5 keV panel only, while
the right scale applies to the other two panels.\label{f4}}
\end{center}
\end{figure}

To investigate the azimuthal distribution of these counts, we show in
Fig.~\ref{f4} the PSFs and observed counts per unit area for the four quadrants
(in three energy bands) with Quad 1 along the proper motion direction (see
Fig.~\ref{f2}).  The shaded region shows the Poisson uncertainty in the surface
brightness measurements.  Here we see significant departures from the PSF in all 
three energy bands. The excesses from $4\arcsec-6\arcsec$ and
$9\arcsec-11\arcsec$ in Quad 2 in the low energy band represents the 
candidate `torus' suggested in \citet{rom03}; no corresponding
excess is seen in Quad 4. A persistent excess is also seen from
$4\arcsec-8\arcsec$ in Quad 3 of the medium energy band, representing
the diffuse emission trailing behind the pulsar.

\begin{figure}[!h]
\begin{center}
\includegraphics[angle=270,scale=0.66]{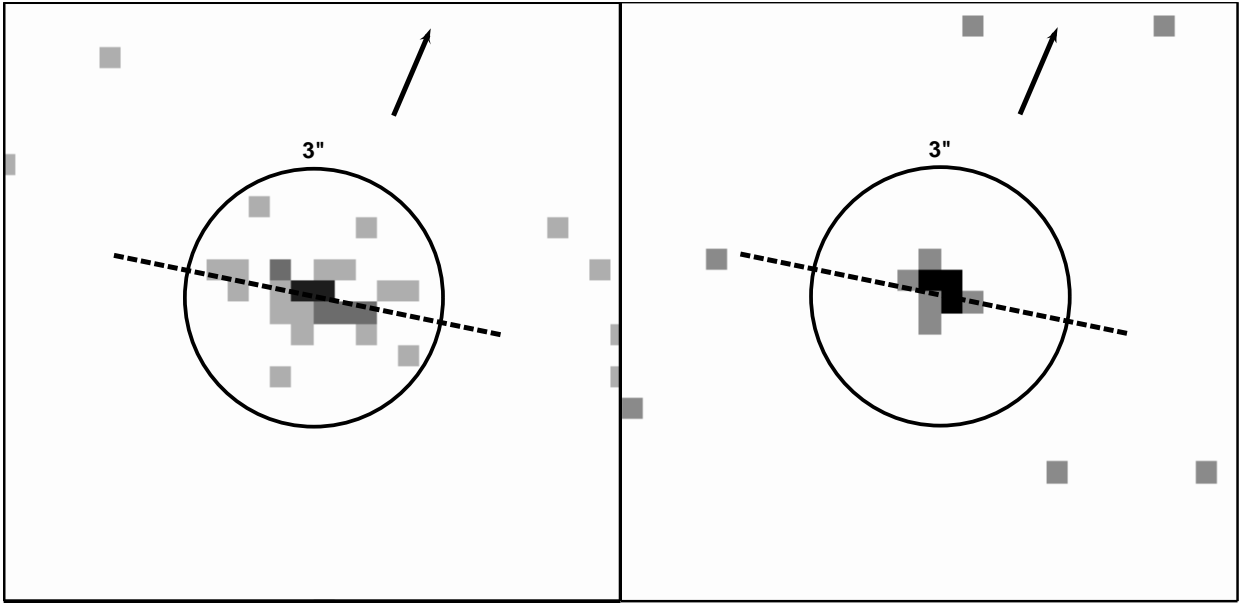}
\caption{\textit{Left:} ACIS-I 4-8 keV image with the dashed line showing
the best-fit PA of the counts. The arrow indicated the pulsar's proper motion
in its LSR. The circle is 3\arcsec\ in radius.
\textit{Right:} Model PSF in the same energy band.\label{f5}}
\end{center}
\end{figure}

	Most intriguing, however, is the very significant excess seen
in the 4-8~keV band from $2\arcsec-4\arcsec$ (Fig.~\ref{f3}).
In Fig.~\ref{f4}
we see that this lies in quadrants 2 \& 4, indicating significant 
structure with a hard spectrum very close to the pulsar. The 4-8~keV ACIS-I 
image (Fig.~\ref{f5}) in fact shows a clear symmetric, almost linear 
structure extending $\sim2\farcs5$ from the pulsar. Realizations of the
4-8~keV model PSF are nearly circular and show no obvious spikes. With 
only 32 counts, we cannot resolve any details or assign a clear origin for 
this emission; however, it should be noted that the X-ray images of
many other young neutron stars show polar jets or equatorial tori. {\it If} 
this is the case for PSR J0538+2817, the symmetry axis could indicate 
the pulsar spin axis.  More importantly, its alignment with the proper 
motion could constrain the pulsar kick physics
(\citealt{rom05}; Ng \& Romani 2007, in preparation).
In order to measure the PA of the symmetry axis
quantitatively, the photon positions were fitted to a straight line
passing through the point source. Linear least squares fitting was employed
and we obtained the best-fit PA at $79\arcdeg\pm4\arcdeg$. 

Although the 4-8\,keV excess is statistically significant, interpretation
of such faint compact structure is of course very uncertain. Some information
on the pulsar orientation may be gleaned from the radio polarization
measurements of \citet{kra03}, which suggest a magnetic inclination 
$\alpha=95\arcdeg$ and impact angle $\sigma=2\arcdeg$, implying a spin 
axis inclination of $\zeta=\alpha+\sigma\approx97\arcdeg$, i.e. very nearly
in the plane of the sky. This means that both the polar jet interpretation
(with the near-orthogonal view implying similar jet/counter-jet fluxes)
and the equatorial torus interpretation (with a thin nearly edge-on torus)
remain viable.  If the extended emission is interpreted as polar jets, 
then the pulsar spin axis is 78\arcdeg\ off the proper motion direction.
This is much larger than any of the other PSR/PWN systems. If the axis is the
equatorial (torus) plane, then the inferred spin vector is at 
PA~$169\arcdeg$($-11\arcdeg$), i.e. 12\arcdeg\ off the velocity vector.

A statistical argument in fact supports the latter interpretation. If
the spin axis and proper motion axis are orthogonal in true space,
then most observed orientations make the projected angle on the plane
of the sky smaller. If they are aligned, the projected angle tends to 
remain nearly aligned. For completely random orientations, 90\% of
vectors separated by $0\arcdeg-10\arcdeg$ retain a 2-D angle $\leq 10\arcdeg$;
for vectors separated by $80\arcdeg-90\arcdeg$ only 29\% retain
$\theta_{2D} \geq 80\arcdeg$. Thus the observed angle is, {\it a priori},
$\sim 3\times$ more likely to arise from an aligned system.
One additional piece of evidence could be extracted from absolute position angle
measurements of the radio polarization \citep[c.f.][]{joh05}. Although mode
ambiguity allows a 90\arcdeg\ jump in the inferred projected field, these authors
note that most pulsars appear to emit in the orthogonal mode, which may allow one
to discriminate between the jet and torus interpretation.

\section{SPECTRAL ANALYSIS}
We now turn to the spectral analysis of the point source. For the best
possible constraints on the spectrum, we have reprocessed both the
18.5~ks cleaned ACIS-S data set and our new 93~ks ACIS-I data with
the latest time-dependent gain calibration and charge transfer
inefficiency (CTI) correction. The source spectrum was extracted
from a 2\arcsec\ radius aperture with the script \texttt{psextract},
and the response matrix files (RMFs) were replaced by the ones built
using the tool \texttt{mkacisrmf}, which accounts for the CTI. To model
the aperture corrections, 10 PSFs with monochromatic energies from 0.5
to 9.5~keV were simulated using ChaRT. The fractional energy encircled
by the aperture as a function of energy is obtained, and then used to
correct the ancillary response files (ARFs). As the ACIS-S data suffer
20\% pile-up, the CCD pile-up model by \citet{dav01} is used in all the
spectral fits.

Results from the combined fits of the ACIS-S and ACIS-I datasets are
listed in Table~\ref{tab2}. All fits are to the 0.3-8~keV range and
the spectral parameter errors reported are projected multidimensional
$1-\sigma$ values. For the uncertainties in flux (and hence stellar radius),
as is often the case with low-statistic CCD-quality data, the projected
errors are very large due to uncertainties in spectral parameters.
Therefore we followed other authors in reporting the single parameter
(i.e. 1-D) $1-\sigma$ error for the flux.

\begin{deluxetable}{cccccccccc}
\tablecaption{SPECTRAL FITS TO PSR J0538+2817\label{tab2}}
\tabletypesize{\scriptsize}
\tablewidth{0pt}
\tablehead{
\colhead{} & \colhead{} & \multicolumn{4}{c}{BLACKBODY/ATMOSPHERE} & \colhead{} & \multicolumn{2}{c}{POWER LAW ($\Gamma=1.5$)} & \colhead{} \\
\cline{3-6} \cline{8-9} \\
\colhead{} & \colhead{} & \colhead{} & \colhead{} & \colhead{Abs. Flux} & \colhead{Unabs. Flux} & \colhead{} & \colhead{Abs. Flux} & \colhead{Unabs. Flux} & \colhead{} \\
\colhead{} & \colhead{} & \colhead{} & \colhead{} & \colhead{$f_{0.5-8}$} & \colhead{$f_{0.5-8}$} & \colhead{} & \colhead{$f_{0.5-8}$} & \colhead{$f_{0.5-8}$} \\
\colhead{} & \colhead{$N_\mathrm{H}$} & \colhead{$T_\infty$} & \colhead{$R_\infty$} & \colhead{($10^{-13}$ ergs} & \colhead{($10^{-13}$ ergs} & \colhead{} & \colhead{($10^{-13}$ ergs} & \colhead{($10^{-13}$ ergs} & \colhead{} \\
\colhead{MODEL} & \colhead{($10^{21}$ cm$^{-2}$)} & \colhead{($10^6\;$K)} & \colhead{(km)} & \colhead{cm$^{-2}$s$^{-1}$)} & \colhead{cm$^{-2}$s$^{-1}$)} & \colhead{} & \colhead{cm$^{-2}$s$^{-1}$)} & \colhead{cm$^{-2}$s$^{-1}$)} & \colhead{$\chi^2$/dof} 
}
\startdata
BB & $2.47^{+0.15}_{-0.14}$ & $2.11^{+0.03}_{-0.04}$ & $2.19\pm0.01$ & $7.24\pm0.07$ & $17.4\pm0.2$ & & \nodata & \nodata & 109.7/139 \\

Atm & $2.94^{+0.07}_{-0.06}$ & $1.05\pm0.05$ & $11.16\pm0.02$ & $7.21\pm0.08$ & $21.4\pm0.2$ & & \nodata & \nodata & 120.7/139 \\

BB+PL & $2.50\pm0.15$ & $2.10^{+0.04}_{-0.03}$ & $2.23\pm0.01$ & $7.15\pm0.07$ & $17.5\pm0.2$ & & $0.20\pm0.06$ & $0.24\pm0.07$ & 99.5/138 \\

Atm+PL & $2.95\pm0.07$ & $1.06^{+0.06}_{-0.05}$ & $10.99\pm0.02$ & $7.14\pm0.08$ & $21.4\pm0.2$ & & $0.20\pm0.05$ & $0.24\pm0.07$ & 110.2/138
\enddata
\end{deluxetable}

The source spectrum is adequately fitted by an absorbed blackbody.
The best-fit $N_\mathrm{H}=2.47\times10^{21}\;\mathrm{cm}^{-2}$
is lower than the previous \textit{CXO} results, but consistent with
the \textit{XMM-Newton} measurements. Comparison with the DM value of
$39.7\;\mathrm{pc\;cm\;^{-3}}$ gives $n_\mathrm{H}/n_e=24$, which
is relatively large.  At the pulsar distance of 1.47~kpc, the best-fit
spectral parameters give an effective blackbody radius
of $R_\infty^{\mathrm{eff}}=2.19\;$km. This is too small to be reconciled
with the whole stellar surface, but the flux could be hot
$T\sim2\times 10^6\;$K emission from a small fraction of the stellar surface
($\sim 2.7\%$ for an $R_\infty=13.1\;$km star), possibly due to some
heating mechanism such as bombardment of the polar cap regions by 
relativistic particles from the magnetosphere. The thermal
radiation from the neutron star surface could also be described by
atmospheric models. Light-element neutron star atmosphere models,
such as those dominated by hydrogen, have large Wien excesses. This gives a
lower effective temperature and hence larger stellar radius in the fit.
We use here a pure H model with $B=10^{12}\;$G \citep{pav92,zav96},
as the inferred surface magnetic field strength of PSR J0538+2817 is
$7\times 10^{11}\;$G \citep{and96}. During the fit, the mass of the
neutron star is held fixed at $M=1.4M_\sun$ and the normalization constant
is fixed using the pulsar distance. The best-fit surface temperature and
radius are $T_\infty^{\mathrm{eff}}=1.05\times10^6\;$K and $R_\infty=11.2\;$km.
As expected, this model suggests a lower effective temperature covering
a large fraction of the neutron star surface for a canonical radius.
The fit is statistically slightly worse than that of the
blackbody model, but both produce quite acceptable $\chi^2$ values.

\begin{figure}[!h!t]
\plotone{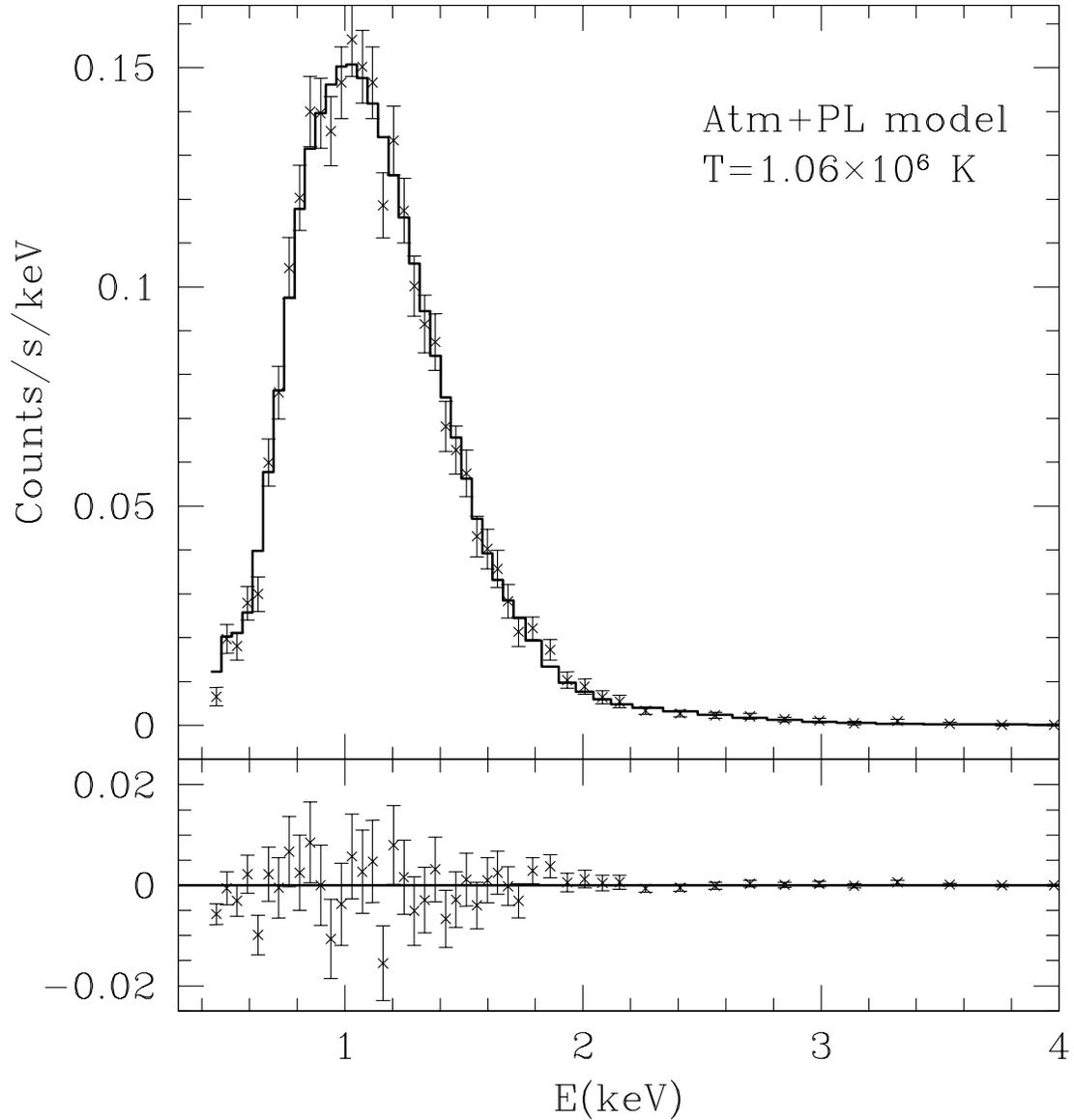}
\caption{Point-source spectrum with a pileup-corrected magnetic H model
atmosphere + power law spectrum and residuals.
\label{f6}}
\end{figure}

The extended emission observed at high energies suggests some flux in the
2\arcsec\ aperture is contributed by non-thermal emission. Therefore we
tried adding a power law component with fixed $\Gamma=1.5$ to the models.
Although the results are not improved substantially, the non-thermal flux
is detected at $>3-\sigma$ level as shown in the table; the effect on
the parameters fitted for the thermal component is very small.
The best-fit atmosphere+power law model is shown in Fig.~\ref{f6}.

\section{DISCUSSION}
\subsection{PSR J0538+2817 / S147 Association}
The association between PSR J0538+2817 and S147 suggested by 
\citet{and96} was based on their positional coincidence and
apparent consistency of the distances and ages. In particular, the 
authors argued
that since the pulsar location is near the SNR center, it is unlikely
to be a chance association. With our accurate proper motion measurement
we can improve this argument.  In order to have a
quantitative estimate of the chance alignment probability, we did simple
Monte Carlo simulations using the model by \citet{fau06}.
Following these authors, we assume the pulsars are born in the Galactic plane
with the galactocentric radial distribution from \citet{yus04} and exponential
distribution in the scale height, and with birth
velocities distributed as a two-component Gaussian model. Acceleration due
to Galactic potential was ignored for simplicity. Our results show that in
$2\times10^5\;$yr, the most extreme age estimate for S147, only
1 pulsar in $2\times10^7$ would have a chance passage within 8\arcmin\  
of the S147 center. We also applied the Maxwellian pulsar velocity distribution
suggested by \citet{hob05}, obtaining nearly identical result.
With a Galactic neutron star birthrate of 2.8 per century 
\citep{fau06} and a radio beaming factor of $\sim1/5$, 
the probability of finding a random, unassociated radio pulsar younger
than 1~Myr which has
passed within 8\arcmin\ of the SNR center is $\sim3\times10^{-4}$. 
This estimate is a conservative upper limit to the probability:
we believe the true age of S147 is considerably younger, and the high 
X-ray temperature of PSR J0538+2817 also implies a younger $\leq 10^5\;$yr age.
Thus a more realistic chance probability is $\geq 10\times$ smaller.
To conclude, PSR J0538+2817 is almost certainly associated with S147 and
this implies a SNR distance of $\sim1.5\;$kpc. This value is
substantially larger than some previous estimates \citep[e.g.][]{kun80},
thus it calls into question some papers that assume a much closer
distance to S147 \citep[e.g.][]{phi81,sal04}.

\subsection{S147 as a Cavity Explosion}

In standard SNR evolution, the shell radius in the Sedov-Taylor
phase is given by $R_{\mathrm{SNR}}=0.31(E_{51}/n_0)^{1/5}t^{2/5}\;$pc
\citep[e.g.][]{swa01}, where the explosion energy is
$E_0=10^{51}E_{51}\;$ergs, age in $t$ years and external medium of
density $n_0\;\mathrm{cm}^{-3}$. The observed angular size of S147
$\theta=83\arcmin$ \citep{sof80} corresponds to a physical
radius of 35~pc at 1.47~kpc. For an age of 30~kyr \citep{kra03},
this requires a very energetic explosion of $E_{51}=20\;n_0$. This suggests
S147 probably occurred in a low density stellar wind bubble, likely 
evacuated by the progenitor star in a WR phase \citep{kra03, gva06}. 
Hence the SNR had a long free-expansion phase, only passing to the
Sedov-Taylor phase when it reached the cavity boundaries at relatively large
radius. This scenario receives further support from
the observed low expansion velocity of S147 at $80\;\mathrm{km\;s^{-1}}$
\citep{kir79}. Note that the progenitor's proper motion can
make the wind bubble asymmetric, with the cavity extending further behind
the star \citep{gva06}. Density gradients in the external medium can also enhance this
asymmetry. Indeed gas and dust surveys suggest that the medium to the South of S147 is denser
and the shell is flattened with the brightest filaments on this side.
Supporting this, the optical observations by \citet{loz76} provides
some hint that the expansion rate is faster in the northern
half of S147. Thus we generally expect the `geometrical center' to lie
somewhat North of the true explosion site.

\subsection{Birth-site of the Progenitor}
With the inferred distance to S147, it is possible to search for the
birth-site of its parent star. O and B stars are the direct progenitors of
neutron stars and the minimum mass for supernova explosion is
$\sim8M_\sun$. These massive stars are generally formed in OB
associations and young open clusters. With their short 
$<50\;$Myr lifetimes they do not travel far from the birth-sites. Typical
peculiar velocities are a few $\mathrm{km\;s^{-1}}$; for $10\;\mathrm{km\;s^{-1}}$, we expect 
the progenitor to travel 
$\leq 500\;$pc. We compiled a list of open clusters and OB associations
from the catalogs including \citet{rup83}, \citet{mel95},
\citet{dia02} \& \citet{kha05} and found only four candidates younger
than 50~Myr with a nominal distance to S147 of $<500\;$pc (Table~\ref{tab3}). 
Directions to potential birth-sites are shown in Fig.~\ref{f1}.

\begin{deluxetable}{llllll}
\tablecaption{Open Clusters younger than 50\,Myr within 500\,pc of S147.\label{tab3}}
\tablewidth{0pt}
\tablehead{
\colhead{Name} & \colhead{d (kpc)} & \colhead{$\theta\:(\arcdeg)$} &
\colhead{r (pc)} & \colhead{N$^*$} & \colhead{Age(Myr)}
}
\startdata
NGC 1960 (M36) &  1.32 & 6.4 & 217 & 60 & 42 \\
OB Gem 1 & 1.34 & 9.9 & 276 & \nodata & $<5$ \\
NGC 2175 & 1.63 & 10.0 & 311 & 60 & 9 \\
Stock 8 & 1.82 & 7.1 & 405 & 40 & 41 \\
NGC 2331 & 1.33 & 19.3 & 488 & 30 & ?
\enddata
\end{deluxetable}

With a 3-D spatial separation of only 220~pc
from S147, \object{NGC 1960} (M36) is the closest candidate\label{s62}. This is
a relatively massive $\sim 40\;$Myr-old cluster \citep{kha05} and 
we consider it the prime candidate birth-site. If correct, the progenitor
traveled from this cluster at PA $\sim175\arcdeg$ with a 3-D space velocity of
$\sim 5(M/10M_\sun)^{2.5}\;\mathrm{km\;s^{-1}}$.
We see no evidence in gas maps or the SNR shell
that the progenitor wind has disturbed the denser cloud to the South. This
and the blowouts of the SNR to the North argue against \object{Gem 1} or
\object{NGC 2175} as the parent cluster. \object{Stock 8} to the North remains viable,
but is distant at $\geq 400\;$pc and less massive. Finally, it is intriguing to note that
the $5^{\rm th}$ nearest young cluster, the very poorly studied \object{NGC 2331},
lies precisely in the direction of the largest (Eastern) extension of the SNR shell;
the H$\alpha$ shell here extends 40\% ($\sim 35^\prime$) further than the
main shock front, with wispy emission present up to $10^\prime$ further
in this direction. It is tempting to associate this blowout with a stellar
wind trail extending along the path to NGC 2331, but this cluster appears
to be relatively low mass and may be too old, with no remaining B stars.

\subsection{Explosion Site}
	From the precise pulsar proper motion, we know that the supernova explosion
must have occurred along the line in Fig.~\ref{f1} with the arrowhead at
the present pulsar position. The explosion site is determined by the true
age of S147/PSR J0538+2817;
tick marks on the line indicate 20, 40 and 60\,kyr ages. The simplest
interpretation is to infer birth at the closest approach to the geometrical
center 36\,kyr ago and assign an uncertainty of $\sim 8\arcmin/57.6\;\mathrm{mas\;yr^{-1}}
= 8\,$kyr. However, if we can define a second axis for the progenitor motion,
we can obtain a more precise age. If we adopt the symmetry
axis suggested by \citet[][solid lines in Fig.~1]{gva06} then the 
birth-site is to the NW of the geometrical center and the intersection 
with the proper motion suggests a SNR age near 20\,kyr. Given the 
rather irregular nature of the Northern
half of the remnant and the argument that the explosion should be
south of the geometrical center, we do not find this axis convincing.
The axis to NGC 2331 intersects at a more plausible explosion age of
30\,kyr.

If however we adopt NGC 1960 as the birth-site, then a explosion somewhat
south of (in front of) the geometrical center becomes natural. Without
a blowout identifying an entry site, we cannot set a precise axis, but
the path should pass close to the geometric center, implying an
intersection with the proper motion vector at $\leq 60\,$kyr.
Our conclusion is that the best estimate of the SNR age is 40\,kyr with
a maximum plausible range of $20-60\,$kyr.  This is slightly older than the 
estimate of \citet{kra03}. However, the inferred
initial spin period of PSR J0538+2817 is not significantly changed. Assuming
magnetic spin-down with constant braking index $n=3$, the initial spin
period is given by
\[ P_0=P\left(1-\frac{n-1}{2}\frac{\tau}{\tau_c}\right ) ^{\frac{1}{n-1}} \;, \]
where $\tau$ and $\tau_c$ are the kinematic and characteristic ages of the
pulsar respectively. We obtained $P_0=138\pm 2.3\;$ms; the kinematic
age of the pulsar is indeed much smaller than its spin-down age.

\subsection{Spin-velocity Alignment}
Only a few pulsars have estimated initial spin
periods; the value for PSR J0538+2817 is the longest among these.
Of course, the high space velocity of the pulsar argues for a strong
birth kick. Note that with the large parallax distance to the pulsar, the
binary break-up scenario described by \citep{gva06} is now even more 
improbable. If one further accepts a progenitor origin in the Galactic 
plane near NGC 1960, then the pulsar's present rapid return to the plane
further supports a birth kick uncorrelated with its parent's motion.

	These considerations make a comparison of this pulsar's kick
and spin direction particularly appealing. Unfortunately the morphology
of the extended structure near the pulsar is not clear enough to define
a definitive spin axis and thus weakens this system's ability to cleanly
test the spin-kick models. However, we can turn the question around: if
the mechanism that seems to cause spin and kick alignment in other
young pulsars acts on PSR J0538+2817, do we expect its spin to be
more nearly aligned or orthogonal? To retain a slow spin with a large
kick velocity, the net kick vector must be nearly radial, applying
little torque to the star. 

	We have performed a series of simulation of neutron star
birth kicks in a range of models where a single thrust is applied
to the surface at fixed angle as the proto-neutron star cools,
with amplitude proportional to the driving neutrino luminosity (Ng
\& Romani 2007, in preparation).
Comparing with the set of all neutron stars with initial spin and/or
speed measurements, we have found (for several models of proto-neutron
star evolution and neutrino cooling) the best-fit parameter distributions
for the neutron star pre-kick spin, the kick amplitude, normal
direction and duration that reproduce the observed pulsar
spin and speed distributions. For these parameters (fixed by a set
of $\sim 50$ other pulsars) we can ask whether a pulsar with
slow initial spin like PSR J0538+2817 is more likely to have its
birth velocity and spin vectors aligned or orthogonal.
The simulations find that
for $P \sim 140\;$ms, the aligned case is produced $30-90\times$ more frequently
than the orthogonal cases, even though faster spin pulsars
do not always show good alignment. Ng \& Romani (2006, in preparation) discuss
the significance of this result for improving kick constraints.

\subsection{Pulsar Thermal Emission}

From the X-ray spectral results, the neutron star atmosphere fit gives
an effective surface temperature of $\gtrsim 10^6\;$K. This matches well to
the standard cooling curve for the pulsar age of $\lesssim 40\;$kyr
\citep[c.f. Fig~6 in][]{mcg03}. Comparing with the cooling
models of \citet{yak04}
we see that our best-fit $1.05\times10^6\;$K surface agrees well with a typical
cooling model at age $25\;$kyr. Even so-called slow cooling neutron
stars (low mass stars, with crustal neutron
pairing and/or accreted low Z envelopes) drop very rapidly below
$T_\infty = 10^6\;$K after $10^5\;$yr. Thus if we interpret the thermal
emission as full surface emission, it seems impossible for 
PSR J0538+2817 to be as old as its characteristic age. In contrast,
its thermal surface emission is quite consistent with its young
$\sim 30\;$kyr kinematic age, requiring no direct Urca
process or any other exotic cooling mechanisms.

{\it XMM} observations find a low 18\% soft X-ray pulsation with a very 
broad profile \citep{mcg03}. These authors interpret this as 
a hot polar cap from a nearly aligned rotator (polar cap axis
and rotation axis both close to the line of sight). This seems at
odds with the radio polarization and PWN data, so a more natural
interpretation might be emission from a gradual temperature variation
across a light element surface, perhaps caused by magnetic dipole
variation in the thermal conductivity \citep{gre83}.

	If interpreted as a re-heated cap emission the thermal flux
would be a surprisingly large 1\% of the full spin-down power. On
the other hand, the non-thermal emission from the PSR/PWN system
as a whole is close to that expected.
In an aperture of radius 15\arcsec, the observed count rate in the
$2-10\;$keV band is $5\times 10^{-3}\;\mathrm{cts\;s^{-1}}$.
After subtracting the background and thermal emission from the pulsar,
the non-thermal contribution gives $2.8\times10^{-3}\;\mathrm{cts\;s^{-1}}$.
For a power law of $\Gamma=1.5$, this converts to the luminosity of
$10^{31.26}\;\mathrm{erg\;s^{-1}}$.
\citet{pos02} found an empirical relation between the X-ray flux
in the 2-10~keV band and spin-down luminosity:
$\log L_{\mathrm{X,(2-10)}}=1.34 \log L_{\mathrm{sd}}-15.34$.
For the case of PSR J0538+2817, the spin-down luminosity derived from the
radio parameters \citep{and96} is
$L_{\mathrm{sd}}=5\times 10^{34}\mathrm{\;ergs\;s^{-1}}$,
which predicts an X-ray luminosity of
$L_{\mathrm{X,(2-10)}}=10^{31.2}\;\mathrm{erg\;s^{-1}}$,
very close to the observed value.

\subsection{Conclusions}
We have reported VLBA astrometric measurements and Chandra ACIS-I
observation of PSR J0538+2817. The VLBA astrometry gives the pulsar
distance of $1.47^{+0.42}_{-0.27}\;$kpc with a precise model independent
transverse velocity $V_\perp=400^{+114}_{-73}\;\mathrm{km\;s^{-1}}$.
These observations strengthen the association with S147 and suggest
NGC 1960 as plausible birth-site for the progenitor star. It seems likely
that the supernova occurred in a stellar wind bubble some 40\,kyr ago.
The X-ray observations of the pulsar show that the thermal point source
has a high temperature consistent with the 40\,kyr age and imply that
it was the source of the observed SNR. Our deep Chandra pointing reveals
extended emission around the pulsar with a very compact symmetric 
structure observed in the 4-8~kev range. The overall PWN flux
is broadly consistent with the emission expected from this relatively
low ${\dot E}$ spin-down pulsar, but the physical origin of the
hard emission in these innermost regions is not clear. Statistical
arguments suggest that the symmetric emission is an equatorial structure,
viewed edge-on, so that the pulsar spin and motion are roughly aligned
as for many other young pulsars. However, further observations are 
needed to reach definitive conclusions.

\acknowledgments
We thank Andrew Lyne for providing the current ephemeris of PSR J0538+2817.
The initial VLBA observations of PSR J0538+2817 were made as part of the
VLBA pulsar astrometry
project (\url{http://www.astro.cornell.edu/~shami/psrvlb/}).
The H$\alpha$ image of S147 is based on data obtained as part of the INT
Photometric H$\alpha$ Survey of the Northern Galactic Plane: prepared by
Albert Zijlstra, University of Manchester and Jonathan Irwin, IoA Cambridge.
The National Radio Astronomy Observatory is a facility of the National Science Foundation operated under cooperative agreement by Associated Universities, Inc.
This work was supported by NASA grant NAG5-13344 and by CXO grant
G05-6058 issued by the Chandra 
X-ray Observatory Center, which is operated by the Smithsonian Astrophysical 
Observatory for and on behalf of the National Aeronautics Space Administration 
under contract NAS8-03060. 

{\it Facilities:} \facility{CXO (ACIS)} \facility{VLBA ()}

\vfill\eject

\end{document}